\documentclass[aps,showpacs,preprintnumbers,amsmath,amssymb]{revtex4}
\usepackage{graphics,epsfig}
\usepackage{graphicx}
\usepackage{dcolumn}
\usepackage{bm}

\begin{document}
\title{{\bf Asymptotic quasinormal modes of a coupled scalar field in the
 Gibbons-Maeda dilaton spacetime}}
\author{Songbai Chen}\email{csb3752@hotmail.com}
\author{Jiliang Jing} \email{jljing@hunnu.edu.cn}
\affiliation{ Institute of Physics and  Department of Physics, \\
Hunan Normal University,  Changsha, Hunan 410081, P. R. China }

\vspace*{0.2cm}
\begin{abstract}
\begin{center}
{\bf Abstract}
\end{center}

Adopting the monodromy technique devised by Motl and Neitzke, we
investigate analytically the asymptotic quasinormal frequencies of
a coupled scalar field in the Gibbons-Maeda dilaton spacetime. We
find that it is described by $ e^{\beta
  \omega}=-[1+2\cos{(\frac{\sqrt{2\xi+1}}{2} \pi)}]-e^{-\beta_I
  \omega}[2+2\cos{(\frac{\sqrt{2\xi+1}}{2}\pi)}]$, which depends on
the structure parameters of the background spacetime and on the
coupling between the scalar and gravitational fields. As the
parameters $\xi$ and $\beta_I$  tend to zero, the real parts of
the asymptotic quasinormal frequencies becomes $T_H\ln{3}$, which
is consistent with Hod's conjecture. When $\xi=\frac{91}{18} $,
the formula becomes that of the Reissner-Nordstr\"{o}m spacetime.
\end{abstract}

 \maketitle
\vspace*{0.2cm}
\section{Introduction}

The study of quasinormal modes in a black hole background
spacetime has a long history \cite{Chandrasekhar}-\cite{Leaver}.
It is shown that their frequencies and damping times are entirely
fixed by the black hole parameters and independent of the initial
perturbations. Therefore, it is certain that quasinormal modes
carry the characteristic information about a black hole and it can
provide a new and direct way for astrophysicists to search black
hole in the universe.

Recently, a great deal of efforts\cite{Hod98}-\cite{jing044} has
been devoted to the study of the asymptotic quasinormal modes
because that Hod's conjecture\cite{Hod98} shows that there maybe
exist a connection between the asymptotic quasinormal frequencies
and quantum gravity. In terms of Bohr's correspondence principle,
the transition frequencies at large quantum numbers ($n\rightarrow
\infty$) should equal to classical oscillation frequencies. Hod
\cite{Hod98} generalized Bohr's correspondence principle to the
black hole physics and  regarded the real parts of the asymptotic
quasinormal frequencies ($n\rightarrow \infty$) as the
characteristic transition frequency for the black hole. Moreover,
Hod observed that the real parts of highly damped quasinormal
frequencies in Schwarzschild black hole can be expressed as
$\omega_R=T_H\ln{3}$, which is derived numerically by
Nollert\cite{Nollert} and later confirmed analytically by Motl
\cite{Motl031}\cite{Motl03} and Andersson \cite{Andersson}.
Together with the first law of black hole thermodynamics, Hod
obtained the value of the fundamental area unit in the
quantization of black hole horizon area. Following Hod's works,
Dreyer\cite{Dreyer03} found that the asymptotic quasinormal modes
can fix the Barbero-Immirzi parameter which is introduced as an
indefinite factor by Immirzi\cite{Immirzi57} to obtain the right
form of the black hole entropy in the loop quantum gravity.
Furthermore, Dreyer obtained that the basic gauge group in the
loop quantum gravity should be $SO(3)$ rather than $SU(2)$. These
exciting new results imply that Hod's conjecture maybe create a
new way to probe the quantum properties of black hole.

However, the question whether Hod's conjecture applies to more
general black holes still remain open. In their deduction, it is
obvious that the factor $\ln{3}$ in the quasinormal frequencies
plays an essential and important role. In other word, whether
Hod's conjecture is valid depends on whether the factor $\ln{3}$
appears in the asymptotic quasinormal frequencies or not.
Recently, we \cite{jing043}
 probed the asymptotic quasinormal modes of a massless scalar field
in the Garfinkle-Horowitz-Strominger dilaton spacetime  and find
that the frequency spectra formula satisfies Hod's conjecture. For
the non-flat spacetime, Cardoso and Yoshida
\cite{Cardoso}\cite{Yoshida}found the asymptotic quasinormal
frequencies in the Schwarzschild de Sitter and Anti-de Sitter
spacetimes depend on the cosmological constant. Only in the case
that the cosmological constant vanishes, the real parts of the
asymptotic quasinormal frequencies returns $T_H\ln{3}$. For the
Reissner-Nordstr\"{o}m black hole, L. Motl and A.
Neitzke\cite{Motl03} found the asymptotic quasinormal frequencies
satisfy
\begin{eqnarray}
e^{\beta \omega}+2+3e^{-\beta_I \omega}=0,\label{01}
\end{eqnarray}
where $\beta$ and $\beta_I$ are the inverse black hole Hawking
temperatures of the outer and inner event horizons respectively.
It is obvious that it is relevant of the electric charge $Q$ and
the real parts of the quasinormal frequencies do not equal to
$T_H\ln{3}$. The more perplexing is that the formula (\ref{01})
does not return the Schwarzschild limit as the black hole charge
$Q$ tends to zero.

Some authors suggested \cite{Berti} that Hod's conjecture about
$T_H \ln{3} $ is valid only in the pure gravitational
perturbations case. The formula (\ref{01}) does not return the
Schwarzschild limit as the black hole charge $Q$ tends to zero
because that it describe a complicated effect of the
material-gravitational coupling. After considering the interaction
between the matter and gravitational fields, Hod's conjecture
maybe possess a more general form. Thus, it is necessary to study
the contribution of the interaction between the matter and
gravitational fields to the asymptotic quasinormal modes in the
more general background spacetimes. Recently, we \cite{jing044}
studied the asymptotic quasinormal modes of a coupled scalar field
in the Garfinkle-Horowitz-Strominger dilaton spacetime. We found
that the asymptotic quasinormal frequencies depend both on the
structure parameters of the background spacetime and on the
coupling between the matter and gravitational fields. Moreover, we
noted that only in the minimal coupled case, the real parts of the
asymptotic quasinormal frequencies is consistent with Hod's
conjecture about $T_H\ln{3}$.

In this paper, our main purpose is to investigate the asymptotic
quasinormal modes of a coupled scalar field in the Gibbons-Maeda
dilaton black hole background. Our plan of the paper is as
follows. In Sec.II, we derive analytically the asymptotic
quasinormal frequency formula of a coupled scalar field in the
Gibbons-Maeda dilaton spacetime by making use of the monodromy
method. At last, a summary and some discussions are presented.

\vspace*{0.2cm}
\section{Asymptotic quasinormal frequencies formula of a coupled scalar field
in the Gibbons-Maeda dilaton spacetime}

The metric for the Gibbons-Maeda dilaton spacetime is
\cite{Gibbons}
\begin{eqnarray}
ds^2=-\frac{(r'-r'_+)(r'-r'_-)}{r'^2-D^2}dt^2+\frac{r'^2-D^2}{(r'-r'_+)(r'-r'_-)}dr'^2
+(r'^2-D^2)(d\theta^2+\sin^2{\theta}d\varphi^2)\label{gem01},
\end{eqnarray}
where $D=(P^2-Q^2)/2M$ and $r'_{\pm}=M\pm \sqrt{M^2+D^2-P^2-Q^2}$.
The parameters $P$ and $Q$ represent the black hole magnetic and
electric charges. When $P=0$ and $D=0$, the metric returns the
Garfinkle-Horowitz-Strominger dilaton spacetime and the
Reissner-Nordstr\"{o}m
 spacetime respectively.

We introduce a coordinate change
\begin{eqnarray}
r^2=r'^2-D^2,
\end{eqnarray}
and the metric (\ref{gem01}) can rewritten as
\begin{eqnarray}
 ds^2&&=-\frac{(\sqrt{r^2+D^2}-r'_+)(\sqrt{r^2+D^2}-r'_-)}{r^2}dt^2\nonumber\\ &&+
 \frac{r^2}{(\sqrt{r^2+D^2}-r'_+)(\sqrt{r^2+D^2}-r'_-)}\frac{r^2}{(r^2+D^2)}dr^2
 +r^2(d\theta^2+\sin^2{\theta}d\varphi^2)\label{gem02}.
\end{eqnarray}
The outer and inner horizons lie in
$r_{\pm}=\sqrt{r'^2_{\pm}-D^2}$.

The general perturbation equation for a coupled massless scalar
field in the dilaton spacetime is \cite{ Frolov}
 \begin{eqnarray}
 \frac{1}{\sqrt{-g}}\partial_\mu(\sqrt{-g}g^{\mu\nu}\partial_\nu)\psi-\xi R\psi=0,\label{eq1}
 \end{eqnarray}
 where $\psi$ is the scalar field and $R$ is the Ricci scalar curvature. The coupling between
 the scalar and gravitational fields is represented by the term $ \xi R\psi$, where
 $\xi$ is a numerical couple factor.

 After adopting WKB approximation $\psi=\frac{e^{-i\omega t}\phi(r)}{r}Y(\theta,\varphi)$,
 introducing the ``tortoise coordinate" change
  \begin{eqnarray}
  x=\sqrt{r^2+D^2}-D+\frac{r'^2_+-D^2}{r'_+-r'_-}
  \ln{\left[\frac{\sqrt{r^2+D^2}-r'_+}{r'_+-D}\right]}-\frac{r'^2_--D^2}{r'_+-r'_-}
  \ln{\left[\frac{\sqrt{r^2+D^2}-r'_-}{r'_--D}\right]},\label{x2}
  \end{eqnarray}
and substituting Eqs.(\ref{gem02}),(\ref{x2}) into Eq.(\ref{eq1}),
we can write the radial perturbation equation for a coupled scalar
  field in the Gibbons-Maeda  dilaton  spacetime as
 \begin{eqnarray}
  \frac{d^2\phi}{dx^2}+(\omega^2-V[r(x)])\phi=0,\label{e3}
 \end{eqnarray}
 where
 \begin{eqnarray}
 &&V[r(x)]=\frac{(\sqrt{r^2+D^2}-r'_+)(\sqrt{r^2+D^2}-r'_-)}{r^2}\times\nonumber\\&&
\left[\frac{l(l+1)}{r^2}-\frac{3D^2(\sqrt{r^2+D^2}-r'_+)(\sqrt{r^2+D^2}-r'_-)-r^2[\sqrt{r^2+D^2}(r'_++r'_-)
 -2r'_+r'_-]}{r^6}+\xi R\right],
 \end{eqnarray}
 and
 \begin{eqnarray}
 R=\frac{2D^2[D^2+r^2-\sqrt{D^2+r^2}(r'_++r'_-)+r'_+r'_-]}{r^6}.
 \end{eqnarray}
As in the Ref.\cite{Motl03}\cite{jing043}, extending analytically
Eq.(\ref{e3}) to the whole complex $r$-plane and comparing the
local with global monodromy of $\phi(r)$ along the selected
contour $L$, we can obtain the asymptotic quasinormal frequency
spectra in the Gibbons-Maeda dilaton black hole spacetime. Because
the regions for the different signs of $Re(x)$ (as shown in Fig.1)
are clearly different from that in the
Garfinkle-Horowitz-Strominger dilaton spacetime case, the contour
$L$ we selected is more complicated than the former. Here, for
avoiding the inner horizon at $r=\sqrt{r'^2_--D^2}$, the selected
contour only goes around the outer horizon at
$r=\sqrt{r'^2_+-D^2}$, where the boundary conditions and the
monodromy are well defined.
\begin{figure}[ht]
\begin{center}
\includegraphics[width=8cm]{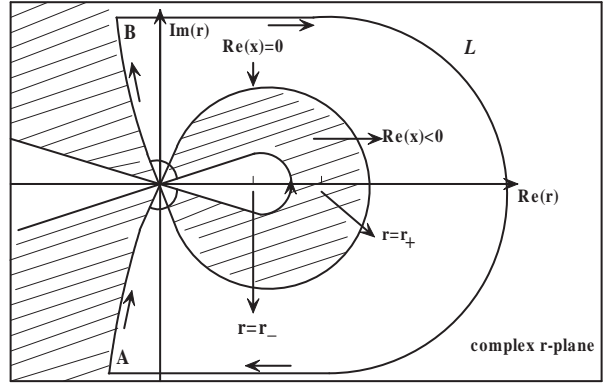}
\caption{The complex $r$-plane and the contour $L$. The regions
with the hachures
 denote the area $Re(x)<0$.}
\end{center}
\label{fig1}
\end{figure}
In order to compute conveniently, we introduce the variable
$z=x-2\pi i\left(\frac{r'^2_+-D^2}{r'_+-r'_-}\right)+2\pi
i\left(\frac{r'^2_--D^2}{r'_+-r'_-}\right)$. For $r=0$, we have
$z=0$. At the vicinity of the point $r=0$, we have
 \begin{eqnarray}
 z\sim \frac{r^4}{4D(r'_+-D)(r'_--D)},\label{z0}
 \end{eqnarray}
 and the behaviors of the Ricci scalar curvature and the potential are
 \begin{eqnarray}
 R\sim \frac{2D^2(r'_+-D)(r'_--D)}{r^6},
 \end{eqnarray}
 and
 \begin{eqnarray}
 V[r(z)]\sim-\frac{3-2\xi}{16z^2}.\label{v4}
 \end{eqnarray}
We make the identification $j=\frac{\sqrt{2\xi+1}}{2}$, and then
the perturbation equation (\ref{e3}) can be rewritten as
 \begin{eqnarray}
 (\frac{d^2}{dz^2}+\omega^2+\frac{1-j^2}{4z^2})\phi(z)=0\label{e4}.
 \end{eqnarray}
 As before, from the boundary condition at point $A$
\begin{eqnarray}
  A_+e^{-i\alpha_+}+A_-e^{-i\alpha_-}=0,
\end{eqnarray}
we easily obtain the
 asymptotic form of $\phi(z)$ at the point marked $A$ is
 \begin{eqnarray}
  \phi(z)\sim (A_+e^{i\alpha_+}+A_-e^{i\alpha_-})e^{-i\omega z}.
\end{eqnarray}
To follow the contour $L$ and approach to point $B$, we first turn
an angle $\frac{\pi}{2}$ around the origin point, corresponding to
$2\pi$ around $z=0$. From the Bessel function behavior near the
origin point, we find that after the $2\pi$ rotation the
asymptotic are
\begin{eqnarray}
 \phi(z)\sim (A_+e^{5i\alpha_+}+A_-e^{5i\alpha_-})e^{-i\omega z}+
 (A_+e^{3i\alpha_+} +A_-e^{3i\alpha_{-}})e^{i\omega z}.
 \label{a02}
 \end{eqnarray}
 Then we must go out along the lobe where the behavior of the wave
 function can be approximated as the purely oscillatory. After
 going around the lobe, we return to the Bessel region a second
 time. However, in this time, we must add an additional distance
 to $z$, namely,
\begin{eqnarray}
 \delta=-2\pi i\left(\frac{r'^2_--D^2}{r'_+-r'_-}\right).
 \end{eqnarray}
Thus, the general solution of the Eq.(\ref{e4}) is
\begin{eqnarray}
 \phi(z)= B_+c_+\sqrt{\omega (z-\delta)}J_{+j/2}(\omega (z-\delta))+
 B_-c_-\sqrt{\omega (z-\delta)}J_{-j/2}(\omega (z-\delta))\label{b01}.
\end{eqnarray}
In terms of the continuity and asymptotic behavior of the wave
function, we have
\begin{eqnarray}
&&A_+e^{3i\alpha_+} +A_-e^{3i\alpha_-}=(B_+e^{i\alpha_+}+
B_-e^{i\alpha_-} )e^{-i\omega \delta},\nonumber\\
&&A_+e^{5i\alpha_+}+A_-e^{5i\alpha_-}=(B_+e^{-i\alpha_+}+B_-e^{-i\alpha_-})e^{i\omega
\delta}.
\end{eqnarray}
To approach the point $B$,  we must finally turn the second $2\pi$
rotation around the point $z=\delta$. The asymptotic forms of the
wave function $\phi(z)$ near the point $B$  becomes
\begin{eqnarray}
 \phi(z)\sim (B_+e^{3i\alpha_+}+B_-e^{3i\alpha_-})e^{i\omega\delta}e^{-i\omega z}+
 (B_+e^{5i\alpha_+} +B_-e^{5i\alpha_{-}})e^{-i\omega\delta}e^{i\omega z}.
 \label{b02}
 \end{eqnarray}
Running over the large semicircle to come back to the point A, we
find that the wave function $e^{-i\omega z}$ must multiply the
factor
\begin{eqnarray}
 \frac{(B_+e^{3i\alpha_+}+B_-e^{3i\alpha_-})e^{i\omega\delta}}
 {A_+e^{i\alpha_+}+A_-e^{i\alpha_-}}=-(1+2\cos{\pi j})-e^{2i\omega
 \delta}(2+2\cos{\pi j}).
 \end{eqnarray}
In the globe monodromy analysis of  $\phi(z)$ around the point
$r=r_+$, we find the coefficient of $e^{-i\omega z}$ must be
multiplied by $e^{\beta \omega}$, where
$\beta=4\pi\left(\frac{r'^2_+-D^2}{r'_+-r'_-}\right)$, is the
inverse black hole Hawking temperature of the outer event horizon.
Then matching the local and globe monodromies of the function
 $\phi(z)$, we find the asymptotic quasinormal frequency spectra
 formula of a coupled scalar field in the Gibbons-Maeda spacetime
 is
 \begin{eqnarray}
  e^{\beta
  \omega}=-[1+2\cos{(\frac{\sqrt{2\xi+1}}{2} \pi)}]-e^{-\beta_I
  \omega}[2+2\cos{(\frac{\sqrt{2\xi+1}}{2} \pi)}],\label{w1}
 \end{eqnarray}
 where $\beta_I=4\pi\left(\frac{r'^2_--D^2}{r'_+-r'_-}\right) $,
 is the inverse black hole Hawking temperature of the inner event horizon.
 As our expectation, the asymptotic quasinormal frequency spectra
 formula of a coupled scalar field in the Gibbons-Maeda dilaton spacetime also depends
 on the coupling between the scalar and the gravitational fields.
 Moreover, we find that as both $P=0$ and $\xi=0$, the real part of
 quasinormal frequency becomes $T_H\ln{3}$, which is consistent with Hod's conjecture.

 \section{summary and discussion}

We have investigated the analytical forms of the asymptotic
quasinormal frequencies for a coupled scalar field in the
Gibbons-Maeda dilaton spacetime by adopting the monodromy
technique. As in the Garfinkle-Horowitz-Strominger dilaton
spacetime, the asymptotic quasinormal frequencies depend not only
on the structure parameters of the background spacetime, but also
on a couple constant $\xi$. The fact tells us again that the
interaction between the matter and gravitational fields affects
the frequencies spectra formula of the asymptotic quasinormal
modes. Moreover, the formulas (\ref{w1}) and (\ref{01}) look like
very similar. Both formulas have a correction term which come from
the phase shift $\delta$. As the formula (\ref{01}) can not yield
the Schwarzschild limit when the electric charge $Q$ tends to
zero, the frequency spectra formula (\ref{w1}) still does not
return to the formula of a coupled scalar field in the
Garfinkle-Horowitz-Strominger dilaton black hole as the magnetic
charge $P$ approaches zero. Similarly, the formula (\ref{w1}) also
can not return that of in the Reissner-Nordstr\"{o}m spacetime as
$D$ approaches to zero. The mathematical reason for these
behaviors is that the variable $z$ and the effective potential $V$
depend heavily on parameters $D$ and $P$. In the cases $D=0$ and
$P=0$, the forms of the variable $z$ and the effective potential
$V$ near the point $z=0$ are clear different from Eqs.(\ref{z0})
and (\ref{v4}). However, when both $\xi$ and $P$ vanish, we find
that the real part of quasinormal frequency in the Gibbons-Maeda
dilaton spacetime can become $T_H\ln{3}$, which agrees with that
of a minimal coupled scalar field in the
Garfinkle-Horowitz-Strominger dilaton spacetime. More
interestingly, when $\xi=\frac{91}{18} $, the formula (\ref{w1})
becomes $e^{\beta \omega}+2+3e^{-\beta_I \omega}=0$, which is
entirely consistent with that of the Reissner-Nordstr\"{o}m black
hole spacetime. This maybe imply that the coupling between the
matter and gravitational field play an important role in the
asymptotic quasinormal modes. The relation between the couple
factor $\xi$ and the asymptotic quasinormal modes in more general
spacetime need to be study more deeply in the future.

\begin{acknowledgments}
This work was supported by the National Natural Science Foundation
of China under Grant No. 10275024; the FANEDD under Grant No.
200317; and the Hunan Provincial Natural Science Foundation of
China under Grant No. 04JJ3019; and the Hunan Normal University
Natural Science Foundation under Grant No.22040639.
\end{acknowledgments}

\end{document}